\documentclass[%
preprint,
 amsmath,amssymb,
 aps,
prstab,
]{revtex4-2}

\usepackage{graphicx}
\usepackage{dcolumn}
\usepackage{bm}

\begin{document}



\title{Generalized Panofsky-Wenzel theorem in curvilinear coordinate systems applicable to non-ultrarelativistic beams}

\author{Demin Zhou}
 \altaffiliation[Also at ]{Graduate University for Advanced Studies (SOKENDAI)}
 \email{dmzhou@post.kek.jp}
\affiliation{%
KEK, High Energy Accelerator Organization, Oho 1-1, Tsukuba 305-0801, Japan
}%
\author{Cheng-Ying Tsai}%
 \email{jcytsai@hust.edu.cn}
 \affiliation{School of Electrical and Electronic Engineering, Huazhong University of Science and Technology, Wuhan 430074, China}

\date{\today}

\begin{abstract}
This note gives an introduction to the theories of impedances and wakes in particle accelerators. The standard formulation assumes that the beam is traveling along a straight orbit with constant velocity $\vec{v}=v\vec{e}_z$. On this note, we show the possibility of extending the formulation for beams traveling along a curved orbit but assuming $|\vec{v}|=v$ to be constant.
\end{abstract}

\keywords{Space charge, impedance}
\maketitle


\section{\label{sec:introduction}Introduction}

To study beam dynamics in particle accelerators, Maxwell's equations, Lorentz force law, and Vlasov (or Vlasov-Fokker-Planck) equation can be taken as the first principles. The evolution of electromagnetic fields and beam distributions is governed by Maxwell's equations and the Vlasov equation, respectively. The two sets of equations are coupled through the Lorentz force. The charged beam guided by external electromagnetic fields also generates fields to its environment. The beam-induced fields act back to the beam and change its distribution. Therefore, the problem to be studied is formulated as a coupled Vlasov-Maxwell equation. With perturbation approximations, the beam motion and the beam-induced fields can be studied separately. The latter is the main topic of this note.

In particle accelerators, the beam-induced fields are usually referred to as wake fields in the literature because they mainly remain behind the source charge at high beam energy. This terminology is followed in this note, but one should note that wake fields can also overtake the source beam in the cases of a beam moving at the velocity of $v<c$, or along a curved trajectory.

To keep the note self-contained, Maxwell's equations and their derivations are introduced in Sec.~\ref{Maxwell_Equations}. The concepts of wake functions and impedances are formulated in a general manner, as shown in Sec.\ref{sec:Fundamental_Formulations}. The Panofsky-Wenzel theorem forms the basis of beam instability theories and is discussed in Sec.~\ref{sec:Panofsky-Wenzel_theorem}. The following sections address possible extensions to the standard theories.

\section{\label{Maxwell_Equations}Maxwell's equations}

Conventionally, the interaction between the beam and the beam-induced fields is described as wake fields and coupling impedance. Therefore, it is natural to start by introducing the fundamental electromagnetic theory. This subsection follows Ref.~\cite{Collin1991} to derive the field equations based on Maxwell's equations. These field equations will build up the basis of this note.

Consider a charged beam traveling with velocity $\vec{v}$ along a prescribed trajectory inside a vacuum chamber. The resulting electromagnetic fields $\vec{E}$ and $\vec{B}$ are governed by Maxwell's equations. In differential form, these equations are
\begin{subequations}
	\begin{equation}
		\nabla \times \vec{E}
		=
		-\frac
		{\partial \vec{B}}
		{\partial t}
		,
		\label{eq:Maxwell-1-1}
	\end{equation}
	\begin{equation}
		\nabla \times \vec{B}
		-
		\mu_0 \epsilon_0
		\frac
		{\partial \vec{E}}
		{\partial t}
		=
		\mu_0
		\vec{J}
		,
		\label{eq:Maxwell-1-2}
	\end{equation}
	\begin{equation}
		\nabla \cdot \vec{B}
		=
		0
		,
		\label{eq:Maxwell-1-3}
	\end{equation}
	\begin{equation}
		\nabla \cdot \vec{E}
		=
		\frac
		{\rho}
		{\epsilon_0}
		,
		\label{eq:Maxwell-1-4}
	\end{equation}
	\begin{equation}
		\nabla \cdot \vec{J}
		=
		-\frac
		{\partial \rho}
		{\partial t}
		.
		\label{eq:Maxwell-1-5}
	\end{equation}
	\label{eq:Maxwell-1}
\end{subequations}
Here, $\rho$ is the charge density, and $\vec{J}$ is the current density. Parameters $\mu_0$ and $\epsilon_0$ are the permeability and permittivity of the free space, respectively. The equation of continuity Eq.~(\ref{eq:Maxwell-1-5}) gives $\vec{J}=\rho \vec{v}$. In the presence of boundaries, extra conditions for fields on the boundaries should be satisfied. Consider a perfectly conducting surface, in general, the boundary conditions take the form of
\begin{equation}
	\vec{n} \times \vec{E}=0,
	\quad
	\vec{n} \cdot \vec{B} = 0,
	\label{eq:GeneralBoundaryConditions1}
\end{equation}
where $\vec{n}$ is the unit vector normal to the surface.

In a vacuum, magnetic induction $\vec{B}$ is proportional to the magnetic field $\vec{H}$ with a simple relation of $\vec{B}=\mu_0\vec{H}$. From Eq.~(\ref{eq:Maxwell-1-3}), the magnetic induction is always solenoidal and may be expressed by the curl of a vector potential $\vec{A}$ as follows:
\begin{equation}
	\vec{B}
	=
	\nabla \times
	\vec{A}
	.
	\label{eq:FluxVector-1}
\end{equation}
Since $\nabla \cdot \nabla \times \vec{A} \equiv 0$, this makes $\nabla \cdot \vec{B}=0$ as well. The vector $\vec{A}$ is called the magnetic vector potential and may have a solenoidal and a lamellar part. At this stage of the analysis, the lamellar part is entirely arbitrary since $\nabla \times \vec{A}_l=0$. Substituting Eq.~(\ref{eq:FluxVector-1}) into the curl equation for $\vec{E}$ gives
\begin{equation}
	\nabla \times
	(
	\vec{E}
	+
	\frac{\partial \vec{A}} {\partial t}
	)
	=
	0
	,
\end{equation}
where $c\equiv \frac{1}{\sqrt{\mu_0\epsilon_0}}$ is the light speed in vacuum. Since $\nabla \times \nabla \Phi \equiv 0$, the above result may be integrated to give
\begin{equation}
	\vec{E}
	=
	-
	\frac{\partial \vec{A}} {\partial t}
	-
	\nabla \Phi
	,
	\label{eq:ElectricFieldFormula1}
\end{equation}
where $\Phi$ is called the electric scalar potential. So far two of Maxwell's equations are satisfied, i.e.\ Eqs.~(\ref{eq:Maxwell-1-1}) and~(\ref{eq:Maxwell-1-3}), and it remains to find the relation between $\Phi$ and $\vec{A}$ and the condition on $\Phi$ and $\vec{A}$ so that the two remaining equations~(\ref{eq:Maxwell-1-2}) and~(\ref{eq:Maxwell-1-4}) are satisfied. The curl equation for $\vec{B}$ gives
\begin{align}
	\nabla \times
	\nabla \times
	\vec{A}
	&=
	\nabla \nabla \cdot \vec{A}
	-
	\nabla^2
	\vec{A}
	\nonumber\\
	&=
	\mu_0 \epsilon_0
	\frac{\partial 
		\vec{E}}
	{\partial t}
	+
	\mu_0
	\vec{J}
	\nonumber\\
	&=
	-
	\frac{1}{c^2}
	\left(
	\frac{\partial^2 \vec{A}}
	{\partial t^2}
	+
	\nabla
	\frac{\partial \Phi} {\partial t}
	\right)
	+
	\mu_0
	\vec{J}
	.
	\label{eq:VectorPotential-tmp1}
\end{align}
Since $\Phi$ and the lamellar part of $\vec{A}$ are as yet arbitrary, one is free to choose a relationship between them. For the purpose of simplification, one can choose
\begin{equation}
	\nabla \cdot \vec{A}
	=
	-
	\frac{1}{c^2}
	\frac{\partial \Phi} {\partial t}
	\label{eq:LorenzGauge}
\end{equation}
which is called the Lorenz gauge condition.
Using Eq.~(\ref{eq:LorenzGauge}), one finds that Eq.~(\ref{eq:VectorPotential-tmp1}) reduces to
\begin{equation}
	\nabla^2 \vec{A}
	-
	\frac{1}{c^2}
	\frac{\partial^2 \vec{A}}
	{\partial t^2}
	=
	-\mu_0
	\vec{J}
	\label{eq:VectorInhomogeneousWaveEquation1}
\end{equation}
Using the Lorenz condition to eliminate $\nabla \cdot \vec{A}$ gives the following equation to be satisfied by the scalar potential $\Phi$:
\begin{equation}
	\nabla^2 \Phi
	-
	\frac{1}{c^2}
	\frac{\partial^2 \Phi} {\partial t^2}
	=
	-
	\frac{\rho}{\epsilon_0}
	.
	\label{eq:ScalarInhomogeneousWaveEquation1}
\end{equation}
Equations~(\ref{eq:VectorInhomogeneousWaveEquation1}) and~(\ref{eq:ScalarInhomogeneousWaveEquation1}) are the vector and scalar inhomogeneous wave equations, respectively. Using the Lorenz condition, the field may be written in terms of the vector potential alone as follows
\begin{subequations}
	\begin{equation}
		\vec{B}
		=
		\nabla \times
		\vec{A}
		,
	\end{equation}
	\begin{equation}
		\vec{E}
		=
		-
		\frac{\partial \vec{A}} {\partial t}
		+
		c^2
		\int^t dt'
		\nabla \nabla \cdot \vec{A}(t')
		.
	\end{equation}
\end{subequations}

Many times, it is more convenient to work with field quantities in the frequency domain rather than in the time domain. Since any physically realizable time-varying function can be decomposed into a spectrum of waves by means of the Fourier integral, there is little loss of generality~\cite{Collin1991}. The Fourier transform may be defined as
\begin{equation}
	F(t)
	=
	\frac{1}{2\pi}
	\int_{-\infty}^{\infty}
	\tilde{F}(\omega)
	e^{-i\omega t}\
	d\omega
	,
	\label{InverseFourierTransformDefinition1}
\end{equation}
and
\begin{equation}
	\tilde{F}(\omega)
	=
	\int_{-\infty}^{\infty}
	F(t)
	e^{i\omega t}\
	dt
	,
	\label{FourierTransformDefinition1}
\end{equation}
where $\omega$ is the radian frequency. The time variation in the form of Eq.~(\ref{InverseFourierTransformDefinition1}) implies that the time differentiations can be replaced by $-i\omega$. In our notation, the field quantities in the frequency domain will be denoted in Roman-type or tilded variables. With the time-varying factor $e^{-i\omega t}$ dropped, the wave equations for potentials change to versions of the inhomogeneous Helmholtz equations
\begin{equation}
	\nabla^2 \vec{\tilde{A}}
	+
	k^2
	\vec{\tilde{A}}
	=
	-\mu_0
	\vec{\tilde{J}}
	\label{eq:VectorHelmholtzEquationFreq1}
\end{equation}
and
\begin{equation}
	\nabla^2 \tilde{\Phi}
	+
	k^2
	\tilde{\Phi}
	=
	-
	\frac{\tilde{\rho}}{\epsilon_0}
	,
	\label{eq:ScalarHelmholtzEquationFreq1}
\end{equation}
where $k\equiv \omega/c$ is the wavenumber. The Lorenz gauge condition Eq.~(\ref{eq:LorenzGauge}) reads
\begin{equation}
	\tilde{\Phi}
	=
	\frac{c^2}{i\omega}
	\nabla \cdot
	\vec{\tilde{A}}
	.
	\label{eq:LorenzGaugeFrequencyDomain1}
\end{equation}
The magnetic induction and electric field are given by
\begin{subequations}
	\begin{equation}
		\vec{\tilde{B}}
		=
		\nabla \times
		\vec{\tilde{A}}
		,
		\label{eq:MagneticInductionFreqDomain1}
	\end{equation}
	\begin{equation}
		\vec{\tilde{E}}
		=
		i\omega \vec{\tilde{A}}
		-
		\nabla \tilde{\Phi}
		=
		i\omega \vec{\tilde{A}}
		-
		\frac{c^2}{i\omega}
		\nabla \nabla \cdot \vec{\tilde{A}}
		.
		\label{eq:ElectricFieldFreqDomain1}
	\end{equation}
\end{subequations}

One can also derive the wave equations of electric fields and magnetic induction directly from the time-domain Maxwell's equations. The equations are
\begin{equation}
	\nabla^2 \vec{E}
	-
	\frac{1}{c^2}
	\frac{\partial^2 \vec{E}}
	{\partial t^2}
	=
	\frac{1}{\epsilon_0}
	\nabla \rho
	+\mu_0
	\frac{\partial \vec{J}}
	{\partial t}
	,
	\label{eq:ElectricFieldInhomogeneousWaveEquation1}
\end{equation}
and
\begin{equation}
	\nabla^2 \vec{B}
	-
	\frac{1}{c^2}
	\frac{\partial^2 \vec{B}}
	{\partial t^2}
	=
	-\mu_0
	\nabla \times
	\vec{J}
	.
	\label{eq:MagneticInductionInhomogeneousWaveEquation1}
\end{equation}
The corresponding equations in the frequency domain are
\begin{equation}
	\nabla^2 \vec{\tilde{E}}
	+
	k^2
	\vec{\tilde{E}}
	=
	\frac{1}{\epsilon_0}
	\nabla \tilde{\rho}
	-i\mu_0kc
	\vec{\tilde{J}}
	,
	\label{eq:ElectricFieldInhomogeneousHelmholtzEquation1}
\end{equation}
and
\begin{equation}
	\nabla^2 \vec{\tilde{B}}
	+
	k^2
	\vec{\tilde{B}}
	=
	-\mu_0
	\nabla \times
	\vec{\tilde{J}}
	.
	\label{eq:MagneticInductionInhomogeneousHelmholtzEquation1}
\end{equation}

The flux density of electromagnetic energy, i.e., Poynting vector, is defined as
\begin{equation}
	\vec{S}
	=
	\frac{1}{\mu_0}
	\vec{E}
	\times
	\vec{B}
	.
	\label{eq:PoyntingVector1}
\end{equation}
In the frequency domain, the Poynting vector is represented by
\begin{equation*}
	\vec{\tilde S} = \frac{1}{{2{\mu _0}}}\vec{\tilde E} \times {\vec{\tilde B}^*}
\end{equation*}

\section{\label{sec:Fundamental_Formulations}Fundamental formulations for impedances and wakes}

In particle accelerators, the charged beam generates electromagnetic fields when traveling inside the vacuum chamber. Beam-induced fields are usually referred to as wake fields in the literature. General formulations of wake fields and their corresponding Fourier transforms (i.e. impedances) in Cartesian coordinates, which are especially suitable for asymmetric structures, have been discussed in the literature (for examples, see Refs.~\cite{Palumbo1994,Heifets:1998er}).

Consider a virtual charged particle $q_0$ moving in parallel to the axis of the vacuum chamber with constant velocity $\vec{v}=\vec{i}_zv$. With the charge density defined by Dirac delta functions as
\begin{equation}
\rho(\vec{R},t)
=
q_0
\delta(x-x_0)
\delta(y-y_0)
\delta(z-z_0)
\label{eq:PointChargeDensity1}
\end{equation}
where $z_0\equiv vt$ and $\vec{R}_0\equiv (x_0,y_0,z_0)$ the spatial vector position, the current density is given by $\vec{J}(\vec{R},t)=\rho(\vec{R},t)\vec{v}$. One can apply them to Maxwell's equations with boundary conditions and obtain the time-varying electromagnetic fields $\vec{E}(\vec{R},t)$ and $\vec{B}(\vec{R},t)$.

\begin{figure}[htbp]
	\centering
	\includegraphics[scale=0.4]{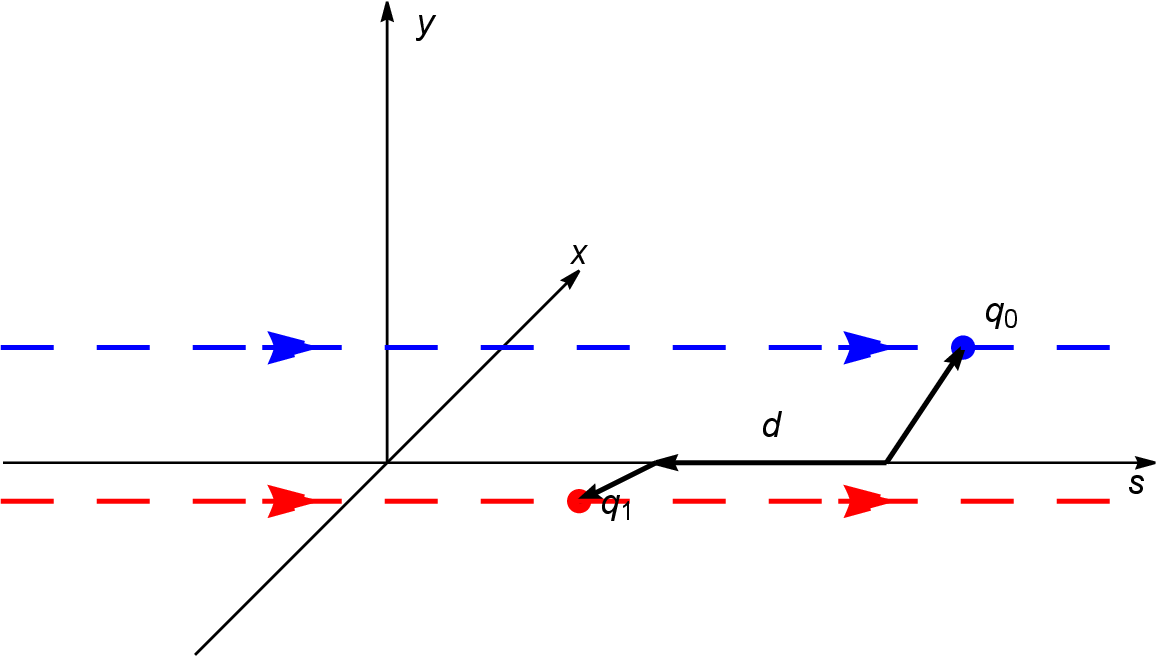}
	\caption[Coordinates of the point charges $q_0$ and $q_1$.]{Coordinates of the point charges $q_0$ and $q_1$. The charge $q_1$ follows $q_0$ for $d>0$ and vice versa for $d<0$.}
	\label{fig:CoordinateSystem1}
\end{figure}

Suppose that a test charged particle $q_1$ with coordinates $\vec{R}=(x,y,z)$ follows $q_0$ with the same velocity $v$ but with a time delay of $\tau = d/v=(z_0-z)/v$ (see Fig.~\ref{fig:CoordinateSystem1}). The Lorentz force acted on $q_1$ is then given by
\begin{equation}
\vec{F}(\vec{R},\vec{R}_0;t)
=
q_1
\left[
\vec{E}(\vec{R},\vec{R}_0;t)
+
\vec{v}
\times
\vec{B}(\vec{R},\vec{R}_0;t)
\right]
.
\label{eq:LorentzForce1}
\end{equation}

To formulate the theory of wake fields and impedance, two approximations are introduced as a basis by following Ref.~\cite{ChaoNotes2002}:
\begin{enumerate}
	\item
	\textbf{The rigid-beam approximation:} This approximation defines the status of the charged beam when it traverses the region considered. It says that the beam is rigid and its motion will not be affected by the wake fields during the traversal of the region. That is, the motions of $q_0$ and $q_1$ in our model will not be affected by the wake fields.
	\item
	\textbf{The impulse approximation:} This approximation defines the effect of the wake fields. It says that the wake effect is only considered as an impulse perturbation applied to the test particle when it completes the traversal; during traversing the region, the wake force will not change the motion of the test particle (rigid-beam approximation). The impulse is defined as the integral of the wake force with respect to the path it traveled.
\end{enumerate}

With the above rigid-beam and impulse approximations, the impulse kick exerted on $q_1$ when it travels through a structure is calculated by integrating the Lorentz force as
\begin{equation}
\overline{\vec{F}}
(\vec{r},\vec{r}_{0};\tau)
=
\int_{-\infty}^{\infty}
dt\
v
\left.
\vec{F}
(\vec{R},\vec{R}_0;t)\right\vert_{z_0=vt,z=vt-d}
.
\quad
\label{eq:WakePotentialDefinition1}
\end{equation}
The vectors $\vec{r}$ and  $\vec{r}_{0}$ in Eq.~(\ref{eq:WakePotentialDefinition1}) represent the transverse positions of the test and source particles respectively, i.e. $\vec{r}=(x,y)$ and $\vec{r}_{0}=(x_0,y_0)$. The integral limits $(-\infty,\infty)$ are chosen here to ensure that the wakefields vanish at the start and end when the beam traverses a structure. In numerical calculations of wakefields, it is more practical to choose finite integral limits. The quantity $\overline{\vec{F}}=(\overline{F}_x,\overline{F}_y,\overline{F}_z)$ is called the wake potential, which is a function of $d$ and the transverse coordinates of the source and test particles. Then, the wake functions in the three dimensions are defined as follows
\begin{subequations}
\begin{equation}
w_z(\vec{r},\vec{r}_{0};d)
=
-
\frac{1}{q_0q_1}
\overline{F}_z
(\vec{r},\vec{r}_{0};\tau)
,
\label{eq:LongitudinalWakeFunctionDefinition1}
\end{equation}
\begin{equation}
w_\perp(\vec{r},\vec{r}_{0};d)
=
\frac{1}{q_0q_1}
\overline{F}_\perp
(\vec{r},\vec{r}_{0};\tau)
,
\label{eq:TransverseWakeFunctionDefinition1}
\end{equation}
\label{eq:WakeFunctionDefinition1}
\end{subequations}
with $d=v\tau$, and $\perp$ indicates $x$ or $y$ for the horizontal and vertical directions, respectively. Here, we use a minus sign for the longitudinal wake function by following the conventional definition of longitudinal wake functions. Using the Fourier transform, one can calculate the spectrum of the wake functions, called impedance, as
\begin{subequations}
\begin{equation}
Z_\parallel(\vec{r},\vec{r}_{0};\omega)
=
\int_{-\infty}^{\infty}
d\tau \
w_z (\vec{r},\vec{r}_{0};d)
e^{i\omega \tau}
,
\label{eq:LongitudinalImpedanceDefinition1}
\end{equation}
\begin{equation}
Z_\perp(\vec{r},\vec{r}_{0};\omega)
=
\kappa
\int_{-\infty}^{\infty}
d\tau \
w_\perp (\vec{r},\vec{r}_{0};d)
e^{i\omega \tau}
,
\label{eq:TransverseImpedanceDefinition1}
\end{equation}
\label{eq:ImpedanceDefinition1}
\end{subequations}
with $\tau=d/v$. Taking the wavenumber of $k\equiv \omega/c$, the impedance is also equivalently written as a function of $k$. The quantity $\kappa$ is defined as a constant to incorporate the convention. Here, we choose $\kappa=i/\beta$ and $\beta=v/c$ the relative velocity by following Ref.~\cite{NgBook2006}. Then, the wake functions expressed by inverting the above Fourier transforms are
\begin{subequations}
\begin{equation}
w_z(\vec{r},\vec{r}_{0};d)
=
\frac{1}{2\pi}
\int_{-\infty}^{\infty}
d\omega \
Z_\parallel (\vec{r},\vec{r}_{0};\omega)
e^{-i\omega \tau}
,
\label{eq:LongitudinalWakeFunctionByImpedance1}
\end{equation}
\begin{equation}
w_\perp(\vec{r},\vec{r}_{0};d)
=
\frac{1}{2\pi\kappa}
\int_{-\infty}^{\infty}
d\omega \
Z_\perp (\vec{r},\vec{r}_{0};\omega)
e^{-i\omega \tau}
.
\label{eq:TransverseWakeFunctionByImpedance1}
\end{equation}
\label{eq:WakeFunctionByImpedance1}
\end{subequations}
The reader may notice that the constant $\kappa$ in Eqs.~(\ref{eq:TransverseImpedanceDefinition1}) and~(\ref{eq:TransverseWakeFunctionByImpedance1}) can be any value. For example, $\kappa=i$ is also often used in the literature.

The above formulation of wake fields and impedance is very general and is applicable to the cases of vacuum chambers with arbitrary shapes. If the chamber considered is cylindrically symmetric, the whole theory can be discussed within the framework of a cylindrical coordinate system. This is relevant to the classical theory, as shown in Ref.~\cite{Alex1993}.

The wake functions and impedance driven by a point charge, as formulated in this section, can be used as Green's functions to calculate the wake potentials and coupling impedance with beam distributions. Let us define $w$ and $W$ as the wake functions of a point charge and a bunch distribution, respectively. Consequently, $Z$ and $\mathcal{Z}$ are the impedance of a point charge and a bunch distribution, respectively. The impedance of a bunch is calculated by integration with respect to the distribution function as
\begin{equation}
	\mathcal{Z}_u(\vec{r};k)
	=
	\int_{-\infty}^{\infty}dx'\int_{-\infty}^{\infty}dy'
	\rho_\perp(x',y')
	Z_u(\vec{r},\vec{r}';k)
	\label{eq:Impedance_Bunch_General1}
\end{equation}
where $u$ represents $x$, $y$, or $\parallel$, and wake functions of a bunch as
\begin{equation}
	W_u(\vec{r};d)
	=
	\int_{-\infty}^{\infty}dx'\int_{-\infty}^{\infty}dy' \int_{-\infty}^{\infty}dz'
	\rho(x',y',z')
	w_u(\vec{r},\vec{r}';d-z')
	\label{eq:Wake_Potential_Bunch_General1}
\end{equation}
where $u$ represents $x$, $y$, or $z$. Here, it is assumed that the bunch distribution is normalized to unity with $\int \rho_\perp(x',y')dx'dy'=1$ and $\int \rho(x',y',z')dx'dy'dz'=1$. Sometimes, the averages of impedance and wake potentials over the transverse density are also useful as an approximation by reducing the model from three-dimensional to one-dimensional:
\begin{equation}
	\overline{\mathcal{Z}}_u(k)
	=
	\int_{-\infty}^{\infty}dx\int_{-\infty}^{\infty}dy
	\rho_\perp(x,y)
	\mathcal{Z}_u(\vec{r};k)
	,
	\label{eq:Impedance_Average_Bunch_General1}
\end{equation}
\begin{equation}
	\overline{W}_u(d)
	=
	\int_{-\infty}^{\infty}dx\int_{-\infty}^{\infty}dy
	\rho_\perp(x,y,d)
	W_u(\vec{r};d)
	.
	\label{eq:Wake_Potential_Average_Bunch_General1}
\end{equation}

\section{\label{sec:Panofsky-Wenzel_theorem}Panofsky-Wenzel theorem}

Using the assumption of constant velocity $\vec{v}=\vec{i}_zv$, from Eq.~(\ref{eq:LorentzForce1}) we have 
\begin{equation}
	\nabla \times \vec{F}(\vec{R},\vec{R}_0;t)
	=
	q_1
	\nabla \times
	\left[
	\vec{E}(\vec{R},\vec{R}_0;t)
	+
	\vec{i}_zv
	\times
	\vec{B}(\vec{R},\vec{R}_0;t)
	\right]
	.
	\label{eq:CurlofLorentzForce1}
\end{equation}
Here $\nabla$ indicates taking the derivative with respect to the coordinates of the test particle $\vec{R}=(x,y,z)$. From the Maxwell's equations, it is trivial to prove
\begin{equation}
	\nabla \times \vec{F}(\vec{R},\vec{R}_0;t)
	=
	-q_1
	\left[
	\frac{\partial}{\partial t}\vec{B}(\vec{R},\vec{R}_0;t)
	+
	v
	\frac{\partial}{\partial z}
	\vec{B}(\vec{R},\vec{R}_0;t)
	\right]
	.
	\label{eq:CurlofLorentzForce2}
\end{equation}

Let $\vec{w}(\vec{r},\vec{r}_{0};d)=(w_x,w_y,w_z)$ with $d=v\tau=z_0-z$ and $z_0=vt$, from the definitions of the wake functions of Eqs.~(\ref{eq:WakeFunctionDefinition1}) we can calculate the curl of $\vec{w}$ as
\begin{equation}
	\nabla' \times \vec{w}(\vec{r},\vec{r}_{0};d)
	=
	\frac{v}{q_0q_1}
	\int_{-\infty}^{\infty}
	dt
	\left[
	\nabla \times
	\vec{F}(\vec{R},\vec{R}_0;t)
	\right]_{z=vt-d}
	.
	\label{eq:CurlofWakeFunction1}
\end{equation}
Here $\nabla'$ indicates taking derivative with respect to $(x,y,d)$. Applying Eq.(\ref{eq:CurlofLorentzForce2}) to Eq.(\ref{eq:CurlofWakeFunction1}) yields
\begin{equation}
	\nabla' \times \vec{w}(\vec{r},\vec{r}_{0};d)
	=
	-\frac{v}{q_0}
	\int_{-\infty}^{\infty}
	dt
	\left[
	\frac{\partial}{\partial t}\vec{B}(\vec{R},\vec{R}_0;t)
	+
	v
	\frac{\partial}{\partial z}
	\vec{B}(\vec{R},\vec{R}_0;t)
	\right]_{z=vt-d}
	.
	\label{eq:CurlofWakeFunction2}
\end{equation}
Here $z_0=vt$ is used. The integral over $t$ leads to
\begin{equation}
	\nabla' \times \vec{w}(\vec{r},\vec{r}_{0};d)
	=
	-\frac{v}{q_0}
	\left.\vec{B}(\vec{R},\vec{R}_0;t)\right\vert_{t=-\infty}^{t={\infty}}
	,
	\label{eq:CurlofWakeFunction3}
\end{equation}
with $z=vt-d$ and $z_0=vt$. When the wakefields vanish at $t=-\infty$ and $t=\infty$, there is 
\begin{equation}
	\left.\vec{B}(\vec{R},\vec{R}_0;t)\right\vert_{t=-\infty}
	=
	\left.\vec{B}(\vec{R},\vec{R}_0;t)\right\vert_{t=\infty}
	.
\end{equation}
It means only the source charge's self-field remains at $t=-\infty$ and $t=\infty$. Finally, we prove that
\begin{equation}
	\nabla' \times \vec{w}(\vec{r}_{1},\vec{r}_{0};d)
	=
	0
	.
	\label{eq:CurlofWakeFunction4}
\end{equation}
This is the so-called Panofsky-Wenzel theorem. Immediately, we can obtain
\begin{equation}
	\frac{\partial w_x}{\partial d}
	=
	\frac{\partial w_z}{\partial x}
	,
	\label{eq:PW-theorem1}
\end{equation}
\begin{equation}
	\frac{\partial w_y}{\partial d}
	=
	\frac{\partial w_z}{\partial y}
	,
	\label{eq:PW-theorem2}
\end{equation}
\begin{equation}
	\frac{\partial w_x}{\partial y}
	=
	\frac{\partial w_y}{\partial x}
	.
	\label{eq:PW-theorem3}
\end{equation}
Equations (\ref{eq:PW-theorem1}) and (\ref{eq:PW-theorem2}) can be rewritten as
\begin{equation}
	\frac{\partial \vec{w}_\perp (\vec{r},\vec{r}_{0};d)}{\partial d}
	=
	\nabla_\perp w_z(\vec{r},\vec{r}_{0};d)
	.
	\label{eq:PW-theorem4}
\end{equation}
Here, $\nabla_\perp$ indicates the derivative with respect to the test particle's transverse coordinates $(x,y)$.

\section{\label{sec:Properties_of_Impedances_and_Wakes}Properties of impedances and wakes}

The wake functions and impedance reflect the fundamental properties of a system and are independent of the charged beam, though they are derived from the response of the system to a point-charge excitation. From the fact that the wake functions are always real, it can be concluded that $Z_\parallel(-\omega)=Z_\parallel^*(\omega)$ and $Z_\perp(-\omega)=-Z_\perp^*(\omega)$. Here, the superscript $^*$ denotes taking the conjugate of a complex number. In many cases, the test particle moving ahead of the source particle does not feel forces; therefore, wake functions are causal, that is, $w(\tau)=0$ if $\tau<0$. This is always true when the charged particle is moving along a straight line with velocity $v=c$ because relativistic causality requires that no signal propagate faster than the speed of light in a vacuum. Causality is a fundamental principle in the physical world. Basically, it states that the effect cannot precede the cause. Here, the causality is introduced in a mathematical way. And it is only for the purpose of convenience in discussing the properties of wake functions. The reader may find that it is not connected to the causality which appears in physical phenomena. There exist various definitions of causality; an interesting discussion can be found in Ref.~\cite{NussenzveigBook1972}.

For causal wake functions, the real and imaginary parts of their impedance are intimately related. The relation can be described based on the Titchmarsh theorem in mathematics (for instance, see Ref.~\cite{NussenzveigBook1972}), which says that the three statements as follows are mathematically equivalent:
\begin{enumerate}
	\item $w(\tau)=0$ if $\tau < 0$ and $w(\tau)$ is a function belonging to the space of the square-integral functions $\bf{L}^2$.
	\item Let $Z(\omega) \in \bf{L}^2$ be the Fourier transform of $w(\tau)$, if $\omega$ is real and if
	\begin{equation}
		Z(\omega)=\lim_{\omega' \to 0} Z(\omega+i\omega')
		,
	\end{equation}
	then $Z(\omega+i\omega')$ is holomorphic in the upper half-plane where $\omega' > 0$. Here ``holomorphic'' means $Z(\omega)$ (The variable $\omega$ is complex.) is complex differentiable at every point $\omega$ in the space under consideration.
	\item Hilbert transforms~\cite{LandauBook1984} connect the real and imaginary parts of $Z(\omega)$ as follows:
	\begin{subequations}
		\begin{equation}
			\hbox{Re}\{Z(\omega)\}
			=
			\frac{1}{\pi}
			\hbox{P.V.}
			\int_{-\infty}^{\infty}
			\frac{\hbox{Im}\{Z(\omega')\}}{\omega'-\omega}d\omega'
			,
		\end{equation}
		\begin{equation}
			\hbox{Im}\{Z(\omega)\}
			=
			-\frac{1}{\pi}
			\hbox{P.V.}
			\int_{-\infty}^{\infty}
			\frac{\hbox{Re}\{Z(\omega')\}}{\omega'-\omega}d\omega'
			,
		\end{equation}
		\label{eq:Kramers-KronigRelations1}
	\end{subequations}
where the symbol $\hbox{P.V.}$ indicates taking the principal value of the relevant integral.
\end{enumerate}
The causality of $W(\tau)$ implies that its Fourier transform $Z(\omega)$ is analytic in the upper complex $\omega$-plane. The real and imaginary parts of $Z(\omega)$ are correlated via the Hilbert transforms. In the literature, Eqs.~(\ref{eq:Kramers-KronigRelations1}) are also called the Kramers-Kronig (K-K) relations~\cite{Kronig1926,Kramers1927}. Alternative forms of K-K relations may be useful for practical calculations. Using the impedance property $Z(-\omega)=Z^*(\omega)$, an alternative by eliminating the negative frequency parts can be derived as follows
\begin{subequations}
	\begin{equation}
		\hbox{Re}\{Z(\omega)\}
		=
		\frac{2}{\pi}
		\hbox{P.V.}
		\int_{0}^{\infty}
		\frac{\omega' \hbox{Im}\{Z(\omega')\}}{\omega'^2-\omega^2}d\omega'
		,
	\end{equation}
	\begin{equation}
		\hbox{Im}\{Z(\omega)\}
		=
		-\frac{2\omega}{\pi}
		\hbox{P.V.}
		\int_{0}^{\infty}
		\frac{\hbox{Re}\{Z(\omega')\}}{\omega'^2-\omega^2}d\omega'
		.
	\end{equation}
	\label{eq:Kramers-KronigRelations2}
\end{subequations}
It is possible to remove the trouble of divergence at $\omega'=\omega$ and improve the convergence of Eqs.~(\ref{eq:Kramers-KronigRelations1}) and~(\ref{eq:Kramers-KronigRelations2}). This results in another alternative as follows
\begin{subequations}
	\begin{equation}
		\hbox{Re}\{Z(\omega)\}
		=
		-\frac{2}{\pi}
		\int_{0}^{\infty}
		\frac{\omega' \left( \hbox{Im}\{Z(\omega')\}-\hbox{Im}\{Z(\omega)\} \right)}{\omega'^2-\omega^2}d\omega'
		,
	\end{equation}
	\begin{equation}
		\hbox{Im}\{Z(\omega)\}
		=
		-\frac{2\omega}{\pi}
		\int_{0}^{\infty}
		\frac{ \hbox{Re}\{Z(\omega')\}-\hbox{Re}\{Z(\omega)\}}{\omega'^2-\omega^2}d\omega'
		.
	\end{equation}
	\label{eq:Kramers-KronigRelations3}
\end{subequations}
It is noteworthy that there is no need to take the principal values of the relevant integrals in the above equations. The K-K relations provide the convenience of determining the imaginary part impedance from the real part, or vice versa. For example, they are very useful when the impedance is calculated using the power spectrum method in electromagnetic theory. In this case, one always calculates real functions of the electromagnetic fields; thus, only real part impedance could be found.

It should be emphasized that the previous discussions only apply to causal wake functions. For noncausal wake functions, such as those of the space charge and coherent synchrotron radiation, the above theories have to be extended. A simple application of the K-K relations may fail to determine the correct answer.

\section{\label{sec:Extended theory of impedances and wakes}Extended theory of impedances and wakes}

Assume that the test particle $q_1$ has a distribution with charge density $\rho_1(\vec{R},\vec{R}_1,t-\tau)$, the Lorentz force Eq.~(\ref{eq:LorentzForce1}) can be equivalently written as
\begin{equation}
	\vec{F}(\vec{R}_1,\vec{R}_0;t)
	=
	\int \int \int
	dV\
	\rho_1(\vec{R},\vec{R}_1,t-\tau)
	\left[
	\vec{E}(\vec{R},\vec{R}_0;t)
	+
	\vec{v}
	\times
	\vec{B}(\vec{R},\vec{R}_0;t)
	\right]
	,
	\label{eq:LorentzForce2}
\end{equation}
where $\vec{R}_1$ denotes the center of the distribution of $q_1$. It is more appropriate to formulate the longitudinal wake potential as the work of the Lorentz force.
\begin{align}
	\overline{F}_\parallel
	(\vec{r},\vec{r}_{0};\tau)
	=
	\int_{-\infty}^{\infty}
	dt\
	\vec{v}
	\cdot
	\vec{F}
	(\vec{R},\vec{R}_0;t)
	.
	\label{eq:LongitudinalWakePotential1}
\end{align}
Substituting Eq.~(\ref{eq:LorentzForce2}) into Eq.~(\ref{eq:LongitudinalWakePotential1}), one gets
\begin{align}
	\overline{F}_\parallel
	(\vec{r}_{1},\vec{r}_{0};\tau)
	=
	\int_{-\infty}^{\infty}
	dt\
	\int \int \int
	dV\
	\rho_1(\vec{R},\vec{R}_1,t-\tau)
	\vec{v}
	\cdot
	\vec{E}(\vec{R},\vec{R}_0;t)
	.
	\label{eq:LongitudinalWakePotential2}
\end{align}
The term of $\rho_1(\vec{R},\vec{R}_1,t-\tau)\vec{v}=\vec{J}_1(\vec{R},\vec{R}_1,t-\tau)$ is recognized to be the current density. Therefore, the longitudinal wake potential can be expressed by
\begin{align}
	\overline{F}_\parallel
	(\vec{r}_{1},\vec{r}_{0};\tau)
	&=
	\int_{-\infty}^{\infty}
	dt\
	\int \int \int
	dV\
	\vec{J}_1(\vec{R},\vec{R}_1,t-\tau)
	\cdot
	\vec{E}(\vec{R},\vec{R}_0;t)
	\nonumber\\
	&
	=
	\int_{-\infty}^{\infty}
	dt'\
	\int \int \int
	dV\
	\vec{J}_1(\vec{R},\vec{R}_1,t')
	\cdot
	\vec{E}(\vec{R},\vec{R}_0;t'+\tau)
	.
	\label{eq:LongitudinalWakePotential3}
\end{align}
The second equality is justified by changing the integration variable $t\to t'+\tau$. Then, the longitudinal wake function reads
\begin{equation}
	w_z(\vec{r}_{1},\vec{r}_{0};\tau)
	=
	-
	\frac{1}{q_0q_1}
	\int_{-\infty}^{\infty}
	dt\
	\int \int \int
	dV\
	\vec{J}_1(\vec{R},\vec{R}_1,t)
	\cdot
	\vec{E}(\vec{R},\vec{R}_0;t+\tau)
	.
	\label{eq:LongitudinalWakeFunctionDefinition2}
\end{equation}
Substituting the Fourier transform of the electric field into the above equation, one can find the longitudinal impedance as follows by comparing it with Eq.~(\ref{eq:LongitudinalWakeFunctionByImpedance1}):
\begin{equation}
	Z_\parallel(\vec{r}_{1},\vec{r}_{0};\omega)
	=
	-
	\frac{1}{q_0q_1}
	\int_{-\infty}^{\infty}
	dt\
	\int \int \int
	dV\
	\vec{J}_1(\vec{R},\vec{R}_1,t)
	\cdot
	\vec{E}(\vec{R},\vec{r}_{0};\omega)
	e^{-i\omega t}
	.
	\label{eq:LongitudinalImpedance2}
\end{equation}
The above equation tells that the longitudinal impedance is obtained once the electric field generated by a beam is found by solving Maxwell's equations in the frequency domain. In particular, for a point charge with constant velocity, i.e., $\vec{J}_1(\vec{R},\vec{R}_1,t)=q_1\delta(\vec{R}-\vec{R}_1) \vec{v}$, one has~\cite{HeifetsKheifets1991}
\begin{equation}
	Z_\parallel(\vec{r}_{1},\vec{r}_{0};\omega)
	=
	-
	\frac{1}{q_0}
	\int_{-\infty}^{\infty}
	dt\
	\vec{v}
	\cdot
	\vec{E}(\vec{R}_1,\vec{r}_{0};\omega)
	e^{-i\omega t}
	,
	\quad
	\text{with}\ z_1=vt.
	\label{eq:LongitudinalImpedance3}
\end{equation}
If one is only interested in the monopole impedance (or wake function), which is usually the dominant term, both the source and test charges can be put on the axis, i.e., $\vec{r}_{1}=\vec{r}_{0}=0$.


In the above discussions, it has been assumed that the beam trajectory is along a straight line. Thus, the point charges are in rectilinear motion. In practice, the direction of beam motion may vary with time. The considered region can also be a free space instead of a vacuum chamber. For instance, in some cases, one must consider a curved trajectory due to external fields inside components such as bending magnets or separators. Then, the above discussions have to be extended in proper ways. One possibility is to choose the local curvilinear coordinate system, and this case is relevant to the case of CSR impedance. If one adopts a Cartesian coordinate system for a curved beam trajectory, both coordinates and the velocity of the beam will vary with time. This case is relevant to the impedance of the undulator radiation.

\section{\label{sec:Generalized_Panofsky-Wenzel_theorem}Generalized Panofsky-Wenzel theorem}

In this section, we discuss the possibility of generalizing the Panofsky-Wenzel theorem. In Eq.(\ref{eq:CurlofLorentzForce1}), one can take $\vec{v}=\vec{v}(t)$ as only a function of time but independent of the spatial coordinates (Actually, this is still from the assumption of rigid-beam approximation.). Consider a Cartesian coordinate system, using the fact that
\begin{equation}
	\nabla \times (\vec{a}\times\vec{b})
	=
	\vec{a}\cdot(\nabla\cdot\vec{b})-\vec{b}(\nabla\cdot\vec{a})
	+(\vec{b}\cdot\nabla)\vec{a}-(\vec{a}\cdot\nabla)\vec{b}
	,
\end{equation} 
we obtain
\begin{equation}
	\nabla \times (\vec{v}\times\vec{B})
	=
	\vec{v}\cdot(\nabla\cdot\vec{B})-\vec{B}(\nabla\cdot\vec{v})
	+(\vec{B}\cdot\nabla)\vec{v}-(\vec{v}\cdot\nabla)\vec{B}
	.
	\label{eq:CurlOfCrossProduct1}
\end{equation}
With $\nabla\cdot\vec{B}=0$ (solenoid law of Maxwell's equations), $\nabla\cdot\vec{v}=0$, and $\nabla\vec{v}=0$ (rigid-beam approximation), the above equation is reduced to
\begin{equation}
	\nabla \times (\vec{v}\times\vec{B})
	=
	-(\vec{v}\cdot\nabla)\vec{B}
	.
\end{equation}
Then Eq.(\ref{eq:CurlofLorentzForce2}) is extended to
\begin{equation}
	\nabla \times \vec{F}(\vec{R},\vec{R}_0;t)
	=
	-q_1
	\left[
	\frac{\partial}{\partial t}\vec{B}(\vec{R},\vec{R}_0;t)
	+
	(\vec{v}\cdot\nabla)
	\vec{B}(\vec{R},\vec{R}_0;t)
	\right]
	.
	\label{eq:CurlofLorentzForce3}
\end{equation}
Note that
\begin{equation}
	\frac{\partial}{\partial t}\vec{B}(\vec{R},\vec{R}_0;t)
	+
	(\vec{v}\cdot\nabla)
	\vec{B}(\vec{R},\vec{R}_0;t)
	=
	\frac{d \vec{B}(\vec{R},\vec{R}_0;t)}{d t}
	.
	\label{eq:Bfielddt1}
\end{equation}
Eventually we obtain the general relation for Lorentz force:
\begin{equation}
	\nabla \times \vec{F}(\vec{R},\vec{R}_0;t)
	=
	-q_1
	\frac{d \vec{B}(\vec{R},\vec{R}_0;t)}{d t}
	.
	\label{eq:CurlofLorentzForce4}
\end{equation}
Performing integration over $t$ to both sides of the above equation, we obtain
\begin{equation}
	\int_{-\infty}^\infty dt \nabla \times \vec{F}(\vec{R},\vec{R}_0;t)
	=
	-q_1
	\left.\vec{B}(\vec{R},\vec{R}_0;t)\right\vert_{t=-\infty}^{t=\infty}
	.
\end{equation}
If the beam induced fields vanish at $t=\pm \infty$ or remain unchanged (i.e. $\vec{B}\vert_{t=-\infty}=\vec{B}\vert_{t=\infty}$), we can obtain
\begin{equation}
	\int_{-\infty}^\infty dt \nabla \times \vec{F}(\vec{R},\vec{R}_0;t)
	=
	0
	.
	\label{eq:Generalized_PW_theorem1}
\end{equation}
This is the generalized Panofsky-Wenzel theorem, which is obtained with the conditions of two approximations as defined in Sec.\ref{sec:Fundamental_Formulations}. Here, the generalization means that we take the assumption that $\vec{v}=\vec{v}(t)$, instead of $\vec{v}$ is constant.

Actually, the divergence of $\vec{F}(\vec{R},\vec{R}_0;t)$ gives another relation of the wake functions~\cite{Vaganian:1995wi,ChaoNotes2002}. From Eq.~(\ref{eq:LorentzForce1}) we have 
\begin{equation}
	\nabla \cdot \vec{F}(\vec{R},\vec{R}_0;t)
	=
	q_1
	\nabla \cdot
	\left[
	\vec{E}(\vec{R},\vec{R}_0;t)
	+
	\vec{v}
	\times
	\vec{B}(\vec{R},\vec{R}_0;t)
	\right]
	.
	\label{eq:DivergenceofLorentzForce1}
\end{equation}
Using the facts of
\begin{equation}
	\nabla \cdot (\vec{a}\times\vec{b})
	=\vec{b}\cdot(\nabla\times\vec{a})
	 -\vec{a}\cdot(\nabla\times\vec{b})
	 ,
\end{equation}
and $\nabla\times\vec{v}=0$ with $\vec{v}=\vec{v}(t)$, Eq.(\ref{eq:DivergenceofLorentzForce1}) is equivalent to
\begin{equation}
	\nabla \cdot \vec{F}(\vec{R},\vec{R}_0;t)
	=
	q_1
	\left[
	\nabla \cdot
	\vec{E}(\vec{R},\vec{R}_0;t)
	-
	\vec{v}
	\cdot
	\left(
	\nabla \times
	\vec{B}(\vec{R},\vec{R}_0;t)
	\right)
	\right]
	.
	\label{eq:DivergenceofLorentzForce2}
\end{equation}
Applying Eqs.(\ref{eq:Maxwell-1-2}) and (\ref{eq:Maxwell-1-4}) to the above equation gives
\begin{equation}
	\nabla \cdot \vec{F}(\vec{R},\vec{R}_0;t)
	=
	q_1
	\left[
	\frac{1}{\gamma^2\epsilon_0}
	\rho
	-
	\frac{1}{c^2}
	\vec{v}\cdot \frac{\partial\vec{E}}{\partial t}
	\right]
	.
	\label{eq:DivergenceofLorentzForce3}
\end{equation}
Similar to Eq.(\ref{eq:Bfielddt1}), there is
\begin{equation}
	\frac{\partial\vec{E}}{\partial t}
	+
	(\vec{v}\cdot\nabla)
	\vec{E}
	=
	\frac{d \vec{E}}{d t}
	.
	\label{eq:Efielddt1}
\end{equation}
Consequently, Eq.(\ref{eq:DivergenceofLorentzForce3}) can be rewritten as
\begin{equation}
	\nabla \cdot \vec{F}(\vec{R},\vec{R}_0;t)
	=
	q_1
	\left[
	\frac{1}{\gamma^2\epsilon_0}
	\rho
	-
	\frac{1}{c^2}
	\vec{v}\cdot
	\left(
	\frac{d }{d t} \vec{E}(\vec{R},\vec{R}_0;t)
	-
	(\vec{v}\cdot\nabla)
	\vec{E}(\vec{R},\vec{R}_0;t)
	\right)
	\right]
	.
	\label{eq:DivergenceofLorentzForce4}
\end{equation}
Performing integration over time to both sides of the above equation, one is to find a new relation of the wake functions in three dimensions.
\begin{equation}
	\int_{-\infty}^\infty dt
	\nabla \cdot \vec{F}(\vec{R},\vec{R}_0;t)
	=
	G(\vec{R},\vec{R}_0)
	.
	\label{eq:DivergenceofLorentzForceIntegrated1}
\end{equation}

\subsection{Source particle $\vec{v}_0(t)=v\vec{e}_z$ in Cartesian system}

This case is relevant to the standard formulation of the Panofsky-Wenzel theorem. With $\vec{v}_0(t)=v\vec{e}_z$ in addition to $d=z_0-z$ and $z_0=vt$, from Eq.(\ref{eq:Generalized_PW_theorem1}) one can derive the relation between the longitudinal and transverse wake functions in a Cartesian system. This is partly discussed in Sec.~\ref{sec:Panofsky-Wenzel_theorem}. Here, we examine the results of Eq.(\ref{eq:DivergenceofLorentzForceIntegrated1}). With the constraints over time and the relative positions of the test and source particles in the longitudinal direction, the independent variables of the function $G$ are reduced to $(\vec{r},\vec{r}_0;d)$:
\begin{equation}
	G(\vec{r},\vec{r}_0;d)
	=
	\frac{q_1}{\gamma^2\epsilon_0}
	\int_{-\infty}^\infty dt \rho
	-
	\left.\frac{q_1v}{c^2} E_z\right\vert_{t=-\infty}^{t=\infty}
	+\frac{q_1v^2}{c^2}
	\int_{-\infty}^\infty dt \left. \frac{\partial}{\partial z}E_z\right\vert_{z=z_0-d}
	.
	\label{eq:DivergenceofLorentzForceIntegrated2}
\end{equation}
The first term on the right side of the above equation is the direct space-charge force. The second term vanishes. For the third term, there is $\frac{\partial}{\partial z}=-\frac{\partial}{\partial d}$, and we obtain
\begin{equation}
	G(\vec{r},\vec{r}_0;d)
	=
	\frac{q_1}{\gamma^2\epsilon_0}
	\int_{-\infty}^\infty dt \rho
	-\frac{q_1v^2}{c^2}
	\frac{\partial}{\partial d}
	\int_{-\infty}^\infty dt \left. E_z\right\vert_{z=z_0-d}
	.
	\label{eq:DivergenceofLorentzForceIntegrated3}
\end{equation}

Similar to Eq.(\ref{eq:CurlofWakeFunction1}), we have
\begin{equation}
	\nabla' \cdot \vec{w}(\vec{r},\vec{r}_{0};d)
	=
	\frac{v}{q_0q_1}
	\int_{\infty}^{\infty}
	dt
	\left[
	\nabla \cdot
	\vec{F}(\vec{R}_1,\vec{R}_0;t)
	\right]_{z=vt-d}
	=
	\frac{v}{q_0q_1}
	G(\vec{r},\vec{r}_0;d)
	.
	\label{eq:DivergenceofWakeFunction1}
\end{equation}
Combining Eqs.(\ref{eq:DivergenceofLorentzForceIntegrated3}) and (\ref{eq:DivergenceofWakeFunction1}), we find the relation
\begin{equation}
	\frac{\partial w_x}{\partial x}
	+
	\frac{\partial w_y}{\partial y}
	+
	\frac{1}{\gamma^2}
	\frac{\partial w_z}{\partial d}
	=
	\frac{v}{\gamma^2q_0\epsilon_0}
	\int_{-\infty}^\infty dt \rho
	.
	\label{eq:DivergenceofWakeFunction2}
\end{equation}
In the ultra-relativistic limit (i.e. $\gamma\rightarrow \infty$), the terms containing $\frac{1}{\gamma^2}$ vanish, resulting in
\begin{equation}
	\frac{\partial w_x}{\partial x}
	+
	\frac{\partial w_y}{\partial y}
	=
	0
	.
	\label{eq:WakeFunctionTransverseRelation1}
\end{equation}
It indicates that the horizontal and vertical wake functions are also correlated.

\subsection{Source particle $\vec{v}_0(t)=v\vec{e}_s$ in Frenet-Serret system}

This case is relevant to the wakefields of coherent synchrotron radiation (CSR). Switching to the Frenet-Serret (F-S) coordinate system is a natural choice. Suppose a reference particle is moving with constant velocity $v$ in the horizontal plane along a curved orbit, which is determined by external static magnetic fields. In the F-S system, with $s=vt$ the path length along the reference orbit, the unit vectors $(\vec{e}_x,\vec{e}_y,\vec{e}_s)$ are $s$ dependent and obey~\cite{AgohThesis}
\begin{subequations}
	\begin{equation}
		\frac{d\vec{e}_x}{ds}
		=
		\frac{1}{\rho_x}\vec{e}_s
		,
		\label{eq:FS_unit_vectors1}
	\end{equation}
	\begin{equation}
		\frac{d\vec{e}_y}{ds}
		=
		0
		,
		\label{eq:FS_unit_vectors2}
	\end{equation}
	\begin{equation}
		\frac{d\vec{e}_s}{ds}
		=
		-\frac{1}{\rho_x}\vec{e}_x
		,
		\label{eq:FS_unit_vectors3}
	\end{equation}
	\label{eq:FS_unit_vectors}
\end{subequations}
where $\rho_x$ is the local bending radius of the reference orbit and it can be $s$-dependent. The velocity of the test particle is given by
\begin{equation}
	\frac{d\vec{r}}{dt}
	=
	\left(\beta_x\vec{e}_x+\beta_y\vec{e}_y+g\vec{e}_s\right)\frac{ds}{dt}
	,
	\label{eq:VelocityInFScoordinate1}
\end{equation}
where $\beta_x=dx/ds$, $\beta_y=dy/ds$, and $g=1+x/\rho_x$. For simplicity, here we take $\beta_x=\beta_y=0$. Consequently, the velocity of the test particle $q_1$ is $\vec{v}=gv\vec{e}_s$. The Lorentz force is decomposed as $\vec{F}=F_x\vec{e}_x+F_y\vec{e}_y+F_s\vec{e}_s$, and in F-S coordinate system Eq.(\ref{eq:LorentzForce1}) gives
\begin{subequations}
	\begin{equation}
		F_x
		=
		q_1\left(E_x-gvB_y\right)
		,
		\label{eq:FS_LorentzForce1}
	\end{equation}
	\begin{equation}
		F_y
		=
		q_1\left(E_y+gvB_x\right)
		,
		\label{eq:FS_LorentzForces2}
	\end{equation}
	\begin{equation}
		F_s
		=
		q_1E_s
		.
		\label{eq:LorentzForce3}
	\end{equation}
	\label{eq:FS_LorentzForce}
\end{subequations}
Similar to the treatment in the Cartesian coordinate system, we define the wake functions as
\begin{subequations}
	\begin{equation}
		w_x(x,y,d)
		=
		\frac{1}{q_0q_1}
		\int_{-\infty}^\infty dtv
		\left.F_x\right\vert_{d=s_0-s}
		,
		\label{eq:FS_WakeFunctionX1}
	\end{equation}
	\begin{equation}
		w_y(x,y,d)
		=
		\frac{1}{q_0q_1}
		\int_{-\infty}^\infty dtv
		\left.F_y\right\vert_{d=s_0-s}
		,
		\label{eq:FS_WakeFunctionY1}
	\end{equation}
	\begin{equation}
		w_z(x,y,d)
		=
		-\frac{1}{q_0q_1}
		\int_{-\infty}^\infty dt
		\left.\vec{v}\cdot \vec{F}\right\vert_{d=s_0-s}
		=
		-\frac{1}{q_0q_1}
		\int_{-\infty}^\infty dt
		\left.gv F_s\right\vert_{d=s_0-s}
		.
		\label{eq:WakeFunctionZ1}
	\end{equation}
	\label{eq:FS_WakeFunction1}
\end{subequations}

In the F-S coordinate system, the divergence and curl of a vector $\vec{A}=A_x\vec{e}_x+A_y\vec{e}_y+A_s\vec{e}_s$ are defined by~\cite{Handbook_AP}
\begin{equation}
	\nabla \cdot\vec{A}
	=
	\frac{1}{g}\left[
	\frac{\partial(gA_x)}{\partial x} + \frac{\partial(gA_y)}{\partial y} + \frac{\partial A_s}{\partial s}
	\right]
	,
	\label{eq:FS_divergence_definition1}
\end{equation}
and
\begin{equation}
	\nabla \times \vec{A}
	=
	\frac{1}{g}\left[
	\frac{\partial(gA_s)}{\partial y} - \frac{\partial A_y}{\partial s}
	\right] \vec{e}_x
	+
	\frac{1}{g}\left[
	\frac{\partial A_x}{\partial s} - \frac{\partial (gA_s)}{\partial x}
    \right] \vec{e}_y
    +
    \left(
    \frac{\partial A_y}{\partial x} - \frac{\partial A_x}{\partial y}
    \right) \vec{e}_s
	.
	\label{eq:FS_curl_definition1}
\end{equation}
With the above definitions, we calculate the curl of Lorentz force. Then, from Eq.(\ref{eq:Generalized_PW_theorem1}), we will try to derive the relation of wake functions in the F-S system. In the F-S system, Eq.(\ref{eq:CurlOfCrossProduct1}) also applies, but the terms need to be redefined:
\begin{equation}
	\nabla \cdot \vec{B}
	=
	\frac{1}{g}\left[
	\frac{\partial(gB_x)}{\partial x} + \frac{\partial(gB_y)}{\partial y} + \frac{\partial B_s}{\partial s}
	\right]
	=0
	,
	\label{eq:CurlOfBfieldInFSsystem1}
\end{equation}
\begin{equation}
	(\vec{B}\cdot \nabla)\vec{v}
	=
	\frac{vB_x}{\rho_x}\vec{e}_s-\frac{vB_s}{\rho_x}\vec{e}_x
	,
\end{equation}
\begin{equation}
	(\vec{v}\cdot\nabla)\vec{B}
	=
	v\frac{\partial}{\partial s}\vec{B}
	=
	\left(
	v\frac{\partial B_x}{\partial s} - \frac{vB_s}{\rho_x}
	\right)\vec{e}_x
	+
	v\frac{\partial B_y}{\partial s}
	\vec{e}_y
	+
	\left(
	v\frac{\partial B_s}{\partial s} + \frac{vB_x}{\rho_x}
	\right)\vec{e}_s.
\end{equation}
Here, it is assumed that $\rho_x$ is constant (i.e., independent of $s$). With $\nabla \cdot \vec{v}=0$, we obtain
\begin{equation}
	\nabla \times (\vec{v}\times \vec{B})
	=
	-v \frac{\partial B_x}{\partial s} \vec{e}_x
	-v \frac{\partial B_y}{\partial s} \vec{e}_y
	-v \frac{\partial B_s}{\partial s} \vec{e}_s
	.
\end{equation}
Finally, the curl of Lorentz force can be written as
\begin{equation}
	\nabla \times \vec{F}
	=
	\frac{1}{g}\left[
	\frac{\partial(gF_s)}{\partial y} - \frac{\partial F_y}{\partial s}
	\right] \vec{e}_x
	+
	\frac{1}{g}\left[
	\frac{\partial F_x}{\partial s} - \frac{\partial (gF_s)}{\partial x}
	\right] \vec{e}_y
	+
	\left(
	\frac{\partial F_y}{\partial x} - \frac{\partial F_x}{\partial y}
	\right) \vec{e}_s
	.
	\label{eq:FS_CurlOfLorentzForce1}
\end{equation}
This equation exactly follows the definition of the curl of a vector in the F-S coordinate system. Then, the last step is to correlate the wake functions with the above relation. Applying the above equation to Eq.(\ref{eq:Generalized_PW_theorem1}), and considering the special case of steady-state CSR, one obtains $\frac{{\partial {F_x}}}{{\partial s}} = \frac{{\partial \left( {g{F_s}} \right)}}{{\partial x}}$. Using Eqs.(\ref{eq:FS_WakeFunction1}), one can find that the relations of wake functions have the same form as Eqs.(\ref{eq:PW-theorem1}), with $d = s_0 -s$
\begin{equation}\label{eq:PW-theorem1x}
\frac{{\partial {w_x}}}{{\partial d}} = \frac{{\partial {w_z}}}{{\partial x}}
\end{equation}

This conclusion is not trivial and needs to be validated.

\subsection{Source particle $\vec{v}_0(t)=v_{z0}(t)\vec{e}_{z0}+v_{x0}(t)\vec{e}_x$ in Cartesian system}

This case is relevant to undulator/wiggler radiation. Taking the bending radius of the reference orbit to be $s$-dependent, i.e. $\rho=\rho(s)$, the formulation in the Frenet-Serret coordinate system, as shown in the previous section, is still valid. Another choice is to consider a Cartesian coordinate system. This can be an exercise for the reader.


\section{Using space charge wakes and impedances as a test}

In this section, we shall use the space charge fields as a test case for the derived generalized Panofsky-Wenzel theorem. Consider a rectangular waveguide with its transverse dimensions in the region of $-a/2<x<a/2$ and $-b/2<y<b/2$. The chamber has an infinite length in the $z$ direction and its walls are assumed to be perfectly conductive. The main task is to find the explicit forms of Eqs.~(\ref{eq:LongitudinalImpedanceDefinition1}) and~(\ref{eq:TransverseImpedanceDefinition1}) in terms of eigenmodes of the rectangular waveguide. We start by solving the inhomogeneous Helmholtz equations for vector and scalar potentials
\begin{equation}
\nabla^2 \vec{A}
+
k^2
\vec{A}
=
-\mu_0
\vec{J}
\label{eq:VectorHelmholtzEquationFreq1}
\end{equation}
and
\begin{equation}
\nabla^2 \Phi
+
k^2
\Phi
=
-
\frac{\rho}{\epsilon_0}
,
\label{eq:ScalarHelmholtzEquationFreq1}
\end{equation}
in the frequency domain under the Lorenz gauge condition of
$
\Phi
=
\frac{c^2}{i\omega}
\nabla \cdot
\vec{A}
$
.
Here we define $k\equiv \omega/c$, and the quantities $\vec{J}$ and $\rho$ are expressed as
\begin{subequations}
\begin{equation}
\rho(\vec{r},k)
=
\frac{q_0}{\beta c}
\delta(x-x_0)
\delta(y-y_0)
e^{ik z/\beta}
,
\label{eq:PointChargeDensitySpectrum1}
\end{equation}
\begin{equation}
\vec{J}(\vec{r},k)
=
\rho(\vec{r},k)\vec{v}
=
\vec{i}_z
q_0
\delta(x-x_0)
\delta(y-y_0)
e^{ik z/\beta}
,
\label{eq:PointChargeCurrentSpectrum1}
\end{equation}
\label{eq:PointChargeSpectrum1}
\end{subequations}
with $\beta$ the relative velocity. The delta function of transverse coordinates can be expanded into the summation of the eigenmodes of the rectangular waveguide as follows
\begin{equation}
\delta(\vec{r}-\vec{r}_{0})
=
\sum_{m=0}^\infty
\sum_{n=0}^\infty
\phi _{mn\nu}\left(\vec {r}\right)
\phi _{mn\nu}\left(\vec {r}_{0}\right)
,
\label{eq:DiracDeltaFunctionExpansion2}
\end{equation}
where $\nu$ represents $x$, $y$, or $z$. And the complete set of orthonormal eigenfunctions in the $x$, $y$ and $z$ directions are 
\begin{subequations}
\begin{equation}
\phi_{mnx}\left(\vec {r}\right)
=
\frac{2}{\sqrt{\left(1+\delta _{m0}\right)a b}}
C_x(x)
S_y(y)
\label{eq:eigen2a},
\end{equation}
\begin{equation}
\phi _{mny}\left(\vec {r}\right)
=
\frac{2}{\sqrt{\left(1+\delta _{n0}\right)a b}}
S_x(x)
C_y(y)
\label{eq:eigen2b},
\end{equation}
\begin{equation}
\phi _{mnz}\left(\vec {r}\right)
=
\frac{2}{\sqrt{a b}}
S_x(x)
S_y(y)
\label{eq:eigen2c}
,
\end{equation}
\end{subequations}
where $\delta_{m0}$ and $\delta_{n0}$ are Kronecker deltas. Here, we define $C_x(x) \equiv \cos \left(k_x (x+a/2)\right)$, $S_x(x) \equiv \sin \left(k_x (x+a/2)\right)$, $C_y(y) \equiv \cos \left(k_y (y+b/2)\right)$, and $S_y(y) \equiv \sin \left(k_y (y+b/2)\right)$ with the transverse wave numbers $k_x=m\pi/a$ and $k_y=n\pi/b$. 

With the delta functions in Eqs.~(\ref{eq:PointChargeSpectrum1}) substituted by Eq.~(\ref{eq:DiracDeltaFunctionExpansion2}) and then applied to Eq.~(\ref{eq:VectorHelmholtzEquationFreq1}), the vector potential can be formulated as~\cite{ZhouThesis2011}
\begin{equation}
\vec {A}\left(\vec{r},\vec{r}_0;k\right)
=
\mu _0 q_0
\beta^2\gamma^2
\vec{i}_z
\sum _{m,n\geq 0} 
\frac{
   \phi _{mnz}\left(\vec{r}\right)
   \phi _{mnz}\left(\vec{r}_{0}\right)}
{k^2+\beta^2\gamma^2k_c^2}
e^{i k z/\beta}
,
\label{eq:PointChargeVectorPotential4}
\end{equation}
with Lorentz factor $\gamma=1/\sqrt{1-\beta^2}$, and $k_c=\sqrt{k_x^2+k_y^2}$ is the cut-off wave number of rectangular chamber. 

Given the explicit expression of the vector potential, one can obtain the electric and magnetic fields by
\begin{subequations}
	\begin{equation}
	\vec{B}
	=
	\nabla \times
	\vec{A}
	,
	\label{eq:MagneticInductionFreqDomain1}
	\end{equation}
	\begin{equation}
	\vec{E}
	=
	ikc \vec{A}
	-
	\nabla \Phi
	=
	ikc \vec{A}
	-
	\frac{c}{ik}
	\nabla \nabla \cdot \vec{A}
	.
	\label{eq:ElectricFieldFreqDomain1}
	\end{equation}
\end{subequations}

With detailed derivations omitted here, the wake function and impedance per unit length can be calculated 
\begin{subequations}
\begin{equation}
\frac{Z_\parallel(\vec{r}_{1},\vec{r}_{0};k)}{L}
=
\frac{4iZ_0k}{ab}
\sum _{m,n\geq 0} 
\frac{
   \phi' _{mnz}\left(\vec{r}_{1}\right)
   \phi' _{mnz}\left(\vec{r}_{0}\right)}
{k^2+\beta^2\gamma^2k_c^2}
,
\label{eq:PointChargeLongImpedanceInWaveguide1}
\end{equation}
\begin{equation}
\frac{Z_x(\vec{r}_{1},\vec{r}_{0};k)}{L}
=
\frac{-4Z_0\beta\kappa}{ab}
\sum _{m,n\geq 0} 
\frac{
   k_x
   \phi' _{mnx}\left(\vec{r}_{1}\right)
   \phi' _{mnz}\left(\vec{r}_{0}\right)}
{k^2+\beta^2\gamma^2k_c^2}
,
\label{eq:PointChargeXImpedanceInWaveguide1}
\end{equation}
\begin{equation}
\frac{Z_y(\vec{r}_{1},\vec{r}_{0};k)}{L}
=
\frac{-4Z_0\beta \kappa}{ab}
\sum _{m,n\geq 0} 
\frac{
   k_y
   \phi' _{mny}\left(\vec{r}_{1}\right)
   \phi' _{mnz}\left(\vec{r}_{0}\right)}
{k^2+\beta^2\gamma^2k_c^2}
,
\label{eq:PointChargeYImpedanceInWaveguide1}
\end{equation}
\label{eq:PointChargeImpedanceInWaveguide1}
\end{subequations}
where $Z_0=\mu_0 c$ is the impedance of vacuum, and the unnormalized eigenfunctions are defined as
\begin{subequations}
\begin{equation}
\phi '_{mnx}\left(\vec{r}\right)
=
C_x(x)
S_y(y)	
\label{eq:eigen3a},
\end{equation}
\begin{equation}
\phi '_{mny}\left(\vec{r}\right)
=
S_x(x)
C_y(y)
\label{eq:eigen3b},
\end{equation}
\begin{equation}
\phi '_{mnz}\left(\vec{r}\right)
=
S_x(x)
S_y(y)
\label{eq:eigen3c}.
\end{equation}
\label{eq:engen3}
\end{subequations}
The wake functions corresponding to Eqs.~(\ref{eq:PointChargeImpedanceInWaveguide1}) are
\begin{subequations}
\begin{equation}
\frac{w_z(\vec{r}_{1},\vec{r}_{0};z)}{L}
=
\frac{2Z_0c}{ab}
\text{sgn}(z)
\sum _{m,n\geq 0} 
   \phi' _{mnz}\left(\vec{r}_{1}\right)
   \phi' _{mnz}\left(\vec{r}_{0}\right)
e^{-\gamma k_c |z|}
,
\label{eq:PointChargeLongWakeFunctionInWaveguide1}
\end{equation}
\begin{equation}
\frac{w_x(\vec{r}_{1},\vec{r}_{0};z)}{L}
=
\frac{-2Z_0c}{\gamma ab}
\sum _{m,n\geq 0} 
   \frac{k_x}{k_c}
   \phi' _{mnx}\left(\vec{r}_{1}\right)
   \phi' _{mnz}\left(\vec{r}_{0}\right)
e^{-\gamma k_c |z|}
,
\label{eq:PointChargeXWakeFunctionInWaveguide1}
\end{equation}
\begin{equation}
\frac{w_y(\vec{r}_{1},\vec{r}_{0};z)}{L}
=
\frac{-2Z_0c}{\gamma ab}
\sum _{m,n\geq 0} 
   \frac{k_y}{k_c}
   \phi' _{mny}\left(\vec{r}_{1}\right)
   \phi' _{mnz}\left(\vec{r}_{0}\right)
e^{-\gamma k_c |z|}
,
\label{eq:PointChargeYWakeFunctionInWaveguide1}
\end{equation}
\label{eq:PointChargeWakeFunctionInWaveguide1}
\end{subequations}
where $\text{sgn}(z)$ denotes the sign function. Similar formulations of space-charge wake potentials were presented in Ref.~\cite{Nogales2012evaluation}.

As a test of the derived theorem, it is straightforward to substitute Eqs. (\ref{eq:PointChargeLongWakeFunctionInWaveguide1}) and (\ref{eq:PointChargeXWakeFunctionInWaveguide1}) into Eq. (\ref{eq:PW-theorem1}) and use the fact that
\begin{equation}
\frac{d|z|}{dz} = \frac{z}{|z|} = \text{sgn}(z).
\end{equation}

It then becomes trivial for the remaining two relations Eqs. (\ref{eq:PW-theorem2},\ref{eq:PW-theorem3}), to which we insert Eqs. (\ref{eq:PointChargeWakeFunctionInWaveguide1}). The equivalent forms of the Panofsky-Wenzel theorem in the frequency domain can be derived by inserting Eqs. (\ref{eq:PW-theorem1},\ref{eq:PW-theorem2},\ref{eq:PW-theorem3}) into Eqs. (\ref{eq:ImpedanceDefinition1})
\begin{subequations}
\begin{equation}
\frac{k}{\kappa \beta^2}Z_x = \frac{i}{\beta}\frac{\partial Z_{\parallel}}{\partial x}
\end{equation}
\begin{equation}
\frac{k}{\kappa \beta^2}Z_y = \frac{i}{\beta}\frac{\partial Z_{\parallel}}{\partial y}
\end{equation}
\label{eq:PW-theorem_FD}
\end{subequations}

It is again trivial to show that Eqs. (\ref{eq:PointChargeImpedanceInWaveguide1}) satisfy Eq. (\ref{eq:PW-theorem_FD}).

\section{Summary}
We revisit the fundamental theories of wake fields and impedance in particle accelerators from the first principles. An attempt is shown to extend the standard Panofsky-Wenzel theorem for $\vec{v}=\vec{i}_z v$ to handle the case of $\vec{v}=\vec{v}(t)$ for non-ultrarelativistic beams. The extended theory is tested with the cases of space charge and CSR. For these two cases (and the problems that accompany them), causality is broken. In particular, the beam moves along a curved orbit in the CSR problem.



\nocite{*}

\bibliography{Impedance_and_Wakes}

\begin{thebibliography}{16}%
\makeatletter
\providecommand \@ifxundefined [1]{%
 \@ifx{#1\undefined}
}%
\providecommand \@ifnum [1]{%
 \ifnum #1\expandafter \@firstoftwo
 \else \expandafter \@secondoftwo
 \fi
}%
\providecommand \@ifx [1]{%
 \ifx #1\expandafter \@firstoftwo
 \else \expandafter \@secondoftwo
 \fi
}%
\providecommand \natexlab [1]{#1}%
\providecommand \enquote  [1]{``#1''}%
\providecommand \bibnamefont  [1]{#1}%
\providecommand \bibfnamefont [1]{#1}%
\providecommand \citenamefont [1]{#1}%
\providecommand \href@noop [0]{\@secondoftwo}%
\providecommand \href [0]{\begingroup \@sanitize@url \@href}%
\providecommand \@href[1]{\@@startlink{#1}\@@href}%
\providecommand \@@href[1]{\endgroup#1\@@endlink}%
\providecommand \@sanitize@url [0]{\catcode `\\12\catcode `\$12\catcode
  `\&12\catcode `\#12\catcode `\^12\catcode `\_12\catcode `\%12\relax}%
\providecommand \@@startlink[1]{}%
\providecommand \@@endlink[0]{}%
\providecommand \url  [0]{\begingroup\@sanitize@url \@url }%
\providecommand \@url [1]{\endgroup\@href {#1}{\urlprefix }}%
\providecommand \urlprefix  [0]{URL }%
\providecommand \Eprint [0]{\href }%
\providecommand \doibase [0]{https://doi.org/}%
\providecommand \selectlanguage [0]{\@gobble}%
\providecommand \bibinfo  [0]{\@secondoftwo}%
\providecommand \bibfield  [0]{\@secondoftwo}%
\providecommand \translation [1]{[#1]}%
\providecommand \BibitemOpen [0]{}%
\providecommand \bibitemStop [0]{}%
\providecommand \bibitemNoStop [0]{.\EOS\space}%
\providecommand \EOS [0]{\spacefactor3000\relax}%
\providecommand \BibitemShut  [1]{\csname bibitem#1\endcsname}%
\let\auto@bib@innerbib\@empty
\bibitem [{\citenamefont {Collin}(1991)}]{Collin1991}%
  \BibitemOpen
  \bibfield  {author} {\bibinfo {author} {\bibfnamefont {R.~E.}\ \bibnamefont
  {Collin}},\ }\href@noop {} {\emph {\bibinfo {title} {Field Theory of Guided
  Waves}}}\ (\bibinfo  {publisher} {IEEE},\ \bibinfo {address} {New York},\
  \bibinfo {year} {1991})\BibitemShut {NoStop}%
\bibitem [{\citenamefont {Palumbo}\ \emph {et~al.}(1994)\citenamefont
  {Palumbo}, \citenamefont {Vaccaro},\ and\ \citenamefont
  {Zobov}}]{Palumbo1994}%
  \BibitemOpen
  \bibfield  {author} {\bibinfo {author} {\bibfnamefont {L.}~\bibnamefont
  {Palumbo}}, \bibinfo {author} {\bibfnamefont {V.~G.}\ \bibnamefont
  {Vaccaro}},\ and\ \bibinfo {author} {\bibfnamefont {M.}~\bibnamefont
  {Zobov}},\ }\href@noop {} {\emph {\bibinfo {title} {Wake Fields and
  Impedance}}},\ \bibinfo {type} {Tech. Rep.}\ \bibinfo {number}
  {LNF-94/041(P)}\ (\bibinfo {year} {September, 1994})\ \bibinfo {note} {``CAS
  - CERN Accelerator School: 5th Advanced Accelerator Physics
  Course''}\BibitemShut {NoStop}%
\bibitem [{\citenamefont {Heifets}\ \emph {et~al.}(1998)\citenamefont
  {Heifets}, \citenamefont {Wagner},\ and\ \citenamefont
  {Zotter}}]{Heifets:1998er}%
  \BibitemOpen
  \bibfield  {author} {\bibinfo {author} {\bibfnamefont {S.}~\bibnamefont
  {Heifets}}, \bibinfo {author} {\bibfnamefont {A.}~\bibnamefont {Wagner}},\
  and\ \bibinfo {author} {\bibfnamefont {B.}~\bibnamefont {Zotter}},\
  }\bibfield  {title} {\bibinfo {title} {{Generalized impedances and wakes in
  asymmetric structures}},\ }\href@noop {} {\  (\bibinfo {year}
  {1998})}\BibitemShut {NoStop}%
\bibitem [{\citenamefont {Chao}(2002)}]{ChaoNotes2002}%
  \BibitemOpen
  \bibfield  {author} {\bibinfo {author} {\bibfnamefont {A.}~\bibnamefont
  {Chao}},\ }\href@noop {} {\emph {\bibinfo {title} {Lecture Notes on Topics in
  Accelerator Physics}}}\ (\bibinfo {year} {November 2002})\ \bibinfo {note}
  {``SLAC-PUB-9574"}\BibitemShut {NoStop}%
\bibitem [{\citenamefont {Ng}(2006)}]{NgBook2006}%
  \BibitemOpen
  \bibfield  {author} {\bibinfo {author} {\bibfnamefont {K.~Y.}\ \bibnamefont
  {Ng}},\ }\href {https://doi.org/10.1142/5835} {\emph {\bibinfo {title}
  {{Physics of intensity dependent beam instabilities}}}}\ (\bibinfo
  {publisher} {World Scientific},\ \bibinfo {address} {Hoboken, NJ},\ \bibinfo
  {year} {2006})\BibitemShut {NoStop}%
\bibitem [{\citenamefont {Chao}(1993)}]{Alex1993}%
  \BibitemOpen
  \bibfield  {author} {\bibinfo {author} {\bibfnamefont {A.}~\bibnamefont
  {Chao}},\ }\bibinfo {title} {Physics of collective beam instabilities in high
  energy accelerators}\ (\bibinfo  {publisher} {Wiley},\ \bibinfo {year}
  {1993})\BibitemShut {NoStop}%
\bibitem [{\citenamefont {Nussenzveig}(1972)}]{NussenzveigBook1972}%
  \BibitemOpen
  \bibfield  {author} {\bibinfo {author} {\bibfnamefont {H.~M.}\ \bibnamefont
  {Nussenzveig}},\ }\href@noop {} {\emph {\bibinfo {title} {Causality and
  Dispersion Relations}}}\ (\bibinfo  {publisher} {Academic Press},\ \bibinfo
  {address} {New York and London},\ \bibinfo {year} {1972})\ \bibinfo {note}
  {p. 27}\BibitemShut {NoStop}%
\bibitem [{\citenamefont {Landau}\ \emph {et~al.}(1984)\citenamefont {Landau},
  \citenamefont {Lifshitz},\ and\ \citenamefont {Pitaevskii}}]{LandauBook1984}%
  \BibitemOpen
  \bibfield  {author} {\bibinfo {author} {\bibfnamefont {L.~D.}\ \bibnamefont
  {Landau}}, \bibinfo {author} {\bibfnamefont {E.~M.}\ \bibnamefont
  {Lifshitz}},\ and\ \bibinfo {author} {\bibfnamefont {L.~P.}\ \bibnamefont
  {Pitaevskii}},\ }\href@noop {} {\emph {\bibinfo {title} {Electrodynamics of
  Continuous Media}}}\ (\bibinfo  {publisher} {Pergamon},\ \bibinfo {address}
  {Oxford},\ \bibinfo {year} {1984})\ \bibinfo {note} {2nd edition}\BibitemShut
  {NoStop}%
\bibitem [{\citenamefont {de~L.~Kronig}(1926)}]{Kronig1926}%
  \BibitemOpen
  \bibfield  {author} {\bibinfo {author} {\bibfnamefont {R.}~\bibnamefont
  {de~L.~Kronig}},\ }\bibfield  {title} {\bibinfo {title} {On the theory of
  dispersion of x-rays},\ }\href@noop {} {\bibfield  {journal} {\bibinfo
  {journal} {J. Opt. Soc. Am.}\ }\textbf {\bibinfo {volume} {12}},\ \bibinfo
  {pages} {547} (\bibinfo {year} {1926})}\BibitemShut {NoStop}%
\bibitem [{\citenamefont {Kramers}(1927)}]{Kramers1927}%
  \BibitemOpen
  \bibfield  {author} {\bibinfo {author} {\bibfnamefont {H.~A.}\ \bibnamefont
  {Kramers}},\ }\bibfield  {title} {\bibinfo {title} {La diffusion de la
  lumiere par les atomes},\ }\href@noop {} {\bibfield  {journal} {\bibinfo
  {journal} {Atti Cong. Intern. Fisica}\ }\textbf {\bibinfo {volume} {2}},\
  \bibinfo {pages} {545} (\bibinfo {year} {1927})}\BibitemShut {NoStop}%
\bibitem [{\citenamefont {Heifets}\ and\ \citenamefont
  {Kheifets}(1991)}]{HeifetsKheifets1991}%
  \BibitemOpen
  \bibfield  {author} {\bibinfo {author} {\bibfnamefont {S.~A.}\ \bibnamefont
  {Heifets}}\ and\ \bibinfo {author} {\bibfnamefont {S.~A.}\ \bibnamefont
  {Kheifets}},\ }\bibfield  {title} {\bibinfo {title} {Coupling impedance in
  modern accelerators},\ }\href {https://doi.org/10.1103/RevModPhys.63.631}
  {\bibfield  {journal} {\bibinfo  {journal} {Rev. Mod. Phys.}\ }\textbf
  {\bibinfo {volume} {63}},\ \bibinfo {pages} {631} (\bibinfo {year}
  {1991})}\BibitemShut {NoStop}%
\bibitem [{\citenamefont {Vaganian}\ and\ \citenamefont
  {Henke}(1995)}]{Vaganian:1995wi}%
  \BibitemOpen
  \bibfield  {author} {\bibinfo {author} {\bibfnamefont {S.}~\bibnamefont
  {Vaganian}}\ and\ \bibinfo {author} {\bibfnamefont {H.}~\bibnamefont
  {Henke}},\ }\bibfield  {title} {\bibinfo {title} {{The Panofsky-Wenzel
  theorem and general relations for the wake potential}},\ }\href@noop {}
  {\bibfield  {journal} {\bibinfo  {journal} {Part. Accel.}\ }\textbf {\bibinfo
  {volume} {48}},\ \bibinfo {pages} {239} (\bibinfo {year} {1995})}\BibitemShut
  {NoStop}%
\bibitem [{\citenamefont {Agoh}(2004)}]{AgohThesis}%
  \BibitemOpen
  \bibfield  {author} {\bibinfo {author} {\bibfnamefont {T.}~\bibnamefont
  {Agoh}},\ }\emph {\bibinfo {title} {Dynamics of Coherent Synchrotron
  Radiation by Paraxial Approximation}},\ \href@noop {} {\bibinfo {type}
  {{Ph.D.} thesis}},\ \bibinfo  {school} {University of Tokyo} (\bibinfo {year}
  {2004})\BibitemShut {NoStop}%
\bibitem [{\citenamefont {Chao}\ \emph {et~al.}(2013)\citenamefont {Chao},
  \citenamefont {Mess}, \citenamefont {Tigner},\ and\ \citenamefont
  {Zimmermann}}]{Handbook_AP}%
  \BibitemOpen
  \bibfield  {author} {\bibinfo {author} {\bibfnamefont {A.}~\bibnamefont
  {Chao}}, \bibinfo {author} {\bibfnamefont {K.}~\bibnamefont {Mess}}, \bibinfo
  {author} {\bibfnamefont {M.}~\bibnamefont {Tigner}},\ and\ \bibinfo {author}
  {\bibfnamefont {F.}~\bibnamefont {Zimmermann}},\ }\href@noop {} {\emph
  {\bibinfo {title} {Handbook of Accelerator Physics and Engineering}}}\
  (\bibinfo  {publisher} {World Scientific},\ \bibinfo {address} {Singapore},\
  \bibinfo {year} {2013})\ \bibinfo {note} {2nd edition}\BibitemShut {NoStop}%
\bibitem [{\citenamefont {Zhou}(2011)}]{ZhouThesis2011}%
  \BibitemOpen
  \bibfield  {author} {\bibinfo {author} {\bibfnamefont {D.}~\bibnamefont
  {Zhou}},\ }\emph {\bibinfo {title} {Coherent Synchrotron Radiation and
  Microwave Instability in Electron Storage Rings}},\ \href@noop {} {\bibinfo
  {type} {{Ph.D.} thesis}},\ \bibinfo  {school} {Graduate University for
  Advanced Studies, Japan} (\bibinfo {year} {2011})\BibitemShut {NoStop}%
\bibitem [{\citenamefont {Nogales}\ \emph {et~al.}(2012)\citenamefont
  {Nogales}, \citenamefont {Marini}, \citenamefont {Mart{\'\i}nez},
  \citenamefont {Melc{\'o}n}, \citenamefont {Pereira}, \citenamefont {Esbert},
  \citenamefont {Soto}, \citenamefont {Cogollos},\ and\ \citenamefont
  {Raboso}}]{Nogales2012evaluation}%
  \BibitemOpen
  \bibfield  {author} {\bibinfo {author} {\bibfnamefont {M.~J.}\ \bibnamefont
  {Nogales}}, \bibinfo {author} {\bibfnamefont {S.}~\bibnamefont {Marini}},
  \bibinfo {author} {\bibfnamefont {B.~G.}\ \bibnamefont {Mart{\'\i}nez}},
  \bibinfo {author} {\bibfnamefont {A.~{\'A}.}\ \bibnamefont {Melc{\'o}n}},
  \bibinfo {author} {\bibfnamefont {F.~Q.}\ \bibnamefont {Pereira}}, \bibinfo
  {author} {\bibfnamefont {V.~B.}\ \bibnamefont {Esbert}}, \bibinfo {author}
  {\bibfnamefont {P.}~\bibnamefont {Soto}}, \bibinfo {author} {\bibfnamefont
  {S.}~\bibnamefont {Cogollos}},\ and\ \bibinfo {author} {\bibfnamefont
  {D.}~\bibnamefont {Raboso}},\ }\bibfield  {title} {\bibinfo {title}
  {Evaluation of time domain electromagnetic fields radiated by constant
  velocity moving particles traveling along an arbitrarily shaped cross-section
  waveguide using frequency domain green's functions},\ }\href@noop {}
  {\bibfield  {journal} {\bibinfo  {journal} {Radio Science}\ }\textbf
  {\bibinfo {volume} {47}},\ \bibinfo {pages} {1} (\bibinfo {year}
  {2012})}\BibitemShut {NoStop}%
\end{thebibliography}%

\end{document}